\documentclass[12pt]{iopart}
\begin{document}
\title[Mass Matrix]{Flavour mixing and mass matrices via anticommuting properties}

\author{R Delbourgo\dag}
\address{\dag\ School of Mathematics and Physics, University of Tasmania,
         Private Bag 37 GPO Hobart, AUSTRALIA 7001}

\ead{Bob.Delbourgo@utas.edu.au}

\begin{abstract}
Five anticommuting property coordinates can accommodate all the known
fundamental particles in their three generations plus more. We describe
the points of difference between this scheme and the standard model and
show how flavour mixing arises through a set of expectation values carried
by a {\em single} Higgs superfield.
\end{abstract}

\submitto{\JPA}

\maketitle

\section{Property from anticommuting coordinates} 
The standard model appears to be on a firm footing with many experimental
verifications of particle states, interactions and decays. Yet there is
some uneasiness in the physics community about the plethora of couplings and 
masses (that are tied to 20 or more parameters) and a lack of understanding 
about the `generation problem'. Symmetry schemes which link the three 
families have been proposed aplenty and they attempt to explain the character
of flavour mixing in weak interactions by invoking various relations
between the mixing parameters due to some assumed underlying idea, such
as permutation subsymmetries, texture zeros, seesaw effects, etc. Great efforts 
have gone into justifying each proposal but none stands out and certainly no 
consensus has been reached on the correct description, even though the $V_{UD}$
matrix elements are quite circumscribed today, in contrast to $V_{LN}$. Thus 
one is led to investigate alternative approaches which may be more predictive.

Recently we have proposed a scheme \cite{RD} where five fundamental anticommuting 
(complex) coordinates $\zeta^\mu$ --  appended to the four (real) space-time 
ones $x^m$ -- are linked to something tangible, namely flavour or property.
The framework therefore describes the {\em  where, when and what} of an
event. Fundamental particle properties are composites \cite{DJW,PDJ} corresponding 
to polynomials in $\zeta$ and are of course limited because of property 
anticommutativity. Another nice feature is the effective reduction in 
`dimension' due to the well-known cancellation between fermionic and bosonic 
degrees of freedom. We outlined a general relativistic extension in which 
gravity fell into the $x-x$ sector, gauge fields in the $x-\zeta$ sector and 
the Higgs field in the $\zeta-\zeta$ sector; however we shall not develop that 
side here. Rather we will focus on the particle aspects of the scheme, 
trying comprehend how the weak flavour mixing comes about and 
attempting to fix the mixing parameters with resulting quark and lepton 
masses.

It is worth stressing that our scheme differs in a number of respects
from the standard model and these will be amplified in the next sections.
The main difference is that there are more than 3 generations---a feature of
other models too. However in our scheme the third and fourth coloured 
generations have quarks which are not weak isodoublet but isotriplet, with 
accompanying quarks $X$ having fermion number $F=1/3$ and charge $Q=-4/3$. 
This might be considered as an {\em awful} prediction, but considering how
little one knows about the top quark, it may not be so terrible. Another
feature is the occurrence of colour {\em sextet} down-type quarks $S$. Provided 
the extra particles are heavy in some sense, the $X$ which can decay weakly into 
conventional down quarks, and $S$ which might provide a small admixture to
nucleon states, might impinge on newly discovered narrow (multiquark?)
resonances that fleetingly appear in some experiments. It is these differences 
from the standard picture that add spice to our scheme but also make it 
vulnerable.

The allocation of fermionic states to $\zeta$-polynomials through a superfield
$\Psi$, a function of $x$ and $\zeta$, is studied in some detail in section 2, 
with emphasis on weak isospin assignments of the combinations and their count
supplied in section 3. (However, gory details are relegated to an appendix.) 
Section 4 deals with the Higgs superfield $\Phi$ and its interaction with $\Psi$.
Finally, the last section is an attempt to work out flavour matrix of the
quarks and leptons engendered by eight real vacuum Higgs field components
and a ninth complex component that is natural source of CP-violation.
We then try to do some numerical work with those expectation values in order
to obtain the $V_{UD}$ and $V_{LN}$ mixing matrices; because there are a lot 
of extra states we have to make some (frail) assumptions so as to achieve some 
simplification, and anyone with experience in mixing matrices will know that 
there is obviously quite a lot that could be improved here. But it is a start.

\section{Superfields} 

Our scheme is founded on a set of five complex $\zeta$ for reasons discussed
in an earlier paper. Four is too few and six is far too many, so our 
construction does accord with SU(5) and SO(10) grand unified models -- and 
this is probably no accident. Recapitulating previous work, the assigments of 
these property coordinates for charge and fermion number are:
$Q(\zeta^0,\zeta^1,\zeta^2,\zeta^3,\zeta^4)=(0,1/3,1/3,1/3,-1)$ and
$F(\zeta^0,\zeta^1,\zeta^2,\zeta^3,\zeta^4)=(1,-1/3,-1/3,-1/3,1).$
`Neutrinicity' is tied to $\zeta^0$, `chromicity' to $\zeta^i$ where
$i=1,2,3$ are the three colours, and charged `leptonicity' to $\zeta^4$ in 
building up properties/flavours. These superfields $\Psi$ and $\Phi$ (both 
overall Bose, note) are expanded in odd and even powers of $\zeta,\bar{\zeta}$
respectively. Thus the fermion field components sit in odd sectors and
the bose field components lie in the even sectors of these expansions:
\begin{equation}
 \Phi(x,\zeta,\bar{\zeta})=\sum_{{\rm even}~r+\bar{r}}
 (\bar{\zeta})^{\bar{r}}\phi_{(\bar{r}),(r)}(x)\left(\zeta\right)^r,
\end{equation}
\begin{equation}
\Psi_\alpha(x,\zeta,\bar{\zeta})=\sum_{{\rm odd}~r+\bar{r}}
(\bar{\zeta})^{\bar{r}}\psi_{\alpha(\bar{r}),(r)}(x)\left(\zeta\right)^r.
\end{equation}
Having said that, the behaviours under the Lorentz group of $\Phi$ and 
$\Psi_\alpha$ are very different, so these superfields are to be regarded
as distinct and ought not to be combined in any way. A nice way of depicting
the component fields is to draw up a six-by-six magic square and look at
the entries therein.

There are so many fermionic $\psi$ entries (namely 512) that they invite 
pruning. We do so by tying reflection about the main diagonal to conjugation,
a sensible choice since $\zeta \leftrightarrow \bar{\zeta}$, or
$$\psi^{(c)}_{\alpha(\bar{r}),(r)}=\psi_{\alpha (r)(\bar{r})},\quad
  \phi_{(\bar{r}),(r)}=\phi^\dag_{(r)(\bar{r})},$$
but also we apply another constraint that superfields are selfdual in some 
way, corresponding to reflection about the cross-diagonal. (The dual operation 
$^\times$ leaves quantum numbers intact.) One particular way is to require
$\Psi$ to be anti-selfdual as this exorcises some embarrassing states hiding in 
the square such as $\bar{\zeta}_0\bar{\zeta}_4\zeta^1\zeta^2\zeta^3$ and 
$\bar{\zeta}_4\zeta^0\zeta^1\zeta^2\zeta^3$ which possess $F=-3$ and $Q=2$
respectively. This condition also disposes of a neutrino-like state in the 
upper right corner too, $\zeta^0\zeta^1\zeta^2\zeta^3\zeta^4$, which may be
a mixed blessing.

From hereon we focus on fermions that are colour singlets or triplets
since we want to concenrate on quarks and leptons. Colour sextets do arise
as well but we shall ignore them for now. Placing the up $U,D,L,N$ states in
the magic square produces the following table {\footnote {we have relabelled
components in an earlier article to conform with coming weak isospin 
assignments and have taken the opportunity to correct the $U$ quarks count
from three to four.}}:
\begin{center}
 \begin{tabular}{|l||c|c|c|c|c|c|}  \hline
   $r\backslash\bar{r}$ & 0 & 1 & 2 & 3 & 4 & 5 \\
  \hline \hline
   0 &   & $L_1,N_1,D_5^c$ &  & $L_5^c,D_1,U_1$ & &   \\
   1 &*& &$L_{2,3},N_{2,3},D^c_{3,6,7},U^c_3$& &$L_6^c,D_2,U_2$&   \\
   2 &   &     *     &          & $L_4,N_4,D^c_{4,8},U^c_4$&  & - \\
   3 & * &           &    *     &        &  -   &   \\
   4 &   &     *     &          &   *    &      & - \\
   5 & * &           &    *     &        &   *  &    \\
 \hline 
 \end{tabular}
\end{center}
The full set of superfield components are normalized and listed in the
appendix, from which one may verify the entries above (- means part of the
dual, while * means complex conjugate about the diagonal). Before turning to
the Higgs superfield $\Phi$, we should mention that the adjoint of $\Psi$ has 
to be defined with appropriate sign changes. For example with only one
$\zeta$, if one writes $\Psi_\alpha=\bar{\zeta}\psi_\alpha+\psi^c_\alpha\zeta$, 
then one should take $\bar{\Psi}^\alpha=-\bar{\psi}^\alpha\zeta
+\bar{\zeta}\bar{\psi^c}^\alpha$. This convention ensures that
$\bar{\Psi}\Psi=\bar{\zeta}\zeta(\bar{\psi}\psi+\bar{\psi^c}\psi^c) =
2\bar{\zeta}\zeta\bar{\psi}\psi,$ so that integration over $\zeta,\bar{\zeta}$
yields $2\bar{\psi}\psi$. (Without that sign change there would have been a 
cancellation between the field and charge conjugate contributions.)

There are two more colour triplet quarks of interest; they are the companions
of $U,D$ in the third and fourth generation, namely: $X_3 \sim \bar{\zeta}_0
\bar{\zeta}_1\zeta^4(1+\bar{\zeta}_2\zeta^2\bar{\zeta}_3\zeta^3)/2$ and
$X_4 \sim \bar{\zeta}_0\bar{\zeta}_1\zeta^4(\bar{\zeta}_2\zeta^2+\bar{\zeta}_3
\zeta^3)/2$. More about them anon.

\section{Weak isopin assignments and new particle signatures} 
Just from the way the anticommuting model is constructed, it will not have
escaped the reader that leptonicity coordinates $\zeta^0, \zeta^4$ comprise a 
weak isodoublet, whereas the chromicity coordinates $\zeta^i$ are weak
isosinglets. By the same reasoning, the conjugates $-\bar{\zeta}_4,
\bar{\zeta}_0$ may be expected to form an isodoublet. Quantifying this 
algebraically, define the raising and lowering operators:
\begin{equation}
 T_+= \zeta^0\partial_4 - \bar{\zeta}_4\bar{\partial}^0, \qquad
 T_-= \zeta^4\partial_0 - \bar{\zeta}_0\bar{\partial}^4,
\end{equation}
where $\partial_\mu \equiv \partial/\partial\zeta^\mu$ and
$\bar{\partial}^\mu \equiv \partial/\partial\bar{\zeta}_\mu$ are
independent derivatives because $\zeta$ is complex. Hence the third component
of weak isospin is given by
\begin{equation}
 2T_3=[T_+,T_-]=\zeta^0\partial_0-\zeta^4\partial_4 +
       \bar{\zeta}_4\bar{\partial}^4-\bar{\zeta}_0\bar{\partial}^0.
\end{equation}
It is thereby easy to establish that the polynomial $(\bar{\zeta}_0\zeta^0 +
\bar{\zeta}_4\zeta^4)$ and therefore its square 2$\bar{\zeta}_0\zeta^0
\bar{\zeta}_4\zeta^4$ are weak isosinglets. Correspondingly, we have a
weak isotriplet in the combinations $(-\bar{\zeta}_4\zeta^0,
(\bar{\zeta}_0\zeta^0-\bar{\zeta}_4\zeta^4)/\sqrt{2},\bar{\zeta}_0\zeta^4)$.

With regard to the heavy generations, $U,D$ together with $X$ make up a 
weak {\em isotriplet} --- signifying a break from the standard model. Because 
$X$ carry charge -4/3, they may decay weakly to the $D$ quarks. 
Colour singlet composites of $X_{3,4}$ with traditional quarks $U,D$ 
will produce new meson states $X\bar{U}$ and $X\bar{D}$ having charges -2 
and -1 respectively; we can also envisage baryon states formed as colour singlet
combinations of $X$ with two traditional quarks $U,D$ and one might look out 
for these too.  Nor should it be forgotten that colour sextet quarks with 
$Q=-1/3$ and $F=1/3$ arise, being associated with the property combinations:
$$S^{(kl)}\sim [\epsilon^{ijk}\zeta^l + \epsilon^{ijl}\zeta^k]
\bar{\zeta}_i\bar{\zeta}_j/4.$$
Thus a small part of proton could conceivably be the admixture $u_ku_lS^{(kl)}$, 
and so on. Perhaps combinations such as these may have a connection
to ephemeral narrow width multiquark resonances found in some experiments.

Leaving such speculation aside, we are now in a position to summarise the 
{\em weak isospin multiplets} of quarks and leptons, based on the table and 
the explicit forms listed in the appendix. Firstly the leptons:
$${\rm \bf doublets:}\quad\left( \begin{array}{c} N_{1,2,3,4} \\ 
       L_{1,2,3,4} \end{array} \right);\quad{\rm \bf singlets:} \quad
       L_{5,6}.$$
Secondly the colour triplet quarks:
$${\rm \bf doublets:}\,\left( \begin{array}{c} U_{1,2} \\ 
       D_{1,2} \end{array} \right);\quad{\rm \bf triplets:}\, 
       \left(\begin{array}{c} U_{3,4} \\ D_{3,4} \\ X_{3,4} \end{array}\right);
       \quad {\rm \bf singlets:}\,\,D_{5,6,7,8}.$$
Thirdly, while we are at it, let us also list the vacuum expectation values
of the anti-selfdual Higgs superfield, again referring to the appendix:
$${\rm \bf singlets:}\,{\cal M},{\cal A}+{\cal B},{\cal C},{\cal D},
   {\cal E}+{\cal F},{\cal G};\quad{\rm \bf in~doublet:}{\cal H},{\cal H}^*;\quad
   {\rm \bf in~triplet:} {\cal A}-{\cal B},{\cal E}-{\cal F}. $$ 

\section{Strong interactions, gauge fields and couplings to matter}
Since the strong interactions are associated with colour the generators are
readily expressed in terms of chromicity coordinates, namely we take the traceless
part (to get SU(3)) of 
\begin{equation}
 T^i_j =\zeta^i\partial_j + \bar{\zeta}_j\bar{\partial}^i,
\end{equation}
obeying the usual U(3) commutation relations,
\begin{equation}
[T^i_j, \,T^k_l] = \delta_j^kT^i_l - \delta_l^iT^k_j.
\end{equation}
We thereby construct the matrix ${\rm A}= A^j_i (T^i_j - \delta^i_j T^k_k/3)$
to contain the eight gluon fields $A$ in much the same way that SU(2)$_L$ weak
bosons are connected with the matrix 
$${\rm W} = W^3T_3 + (W^+T_- + W^-T_+)/\sqrt{2},$$
where the weak isospin generators are given in eqns. (3) and (4).

It is quite straightforward to accommodate the standard model components in this
scheme since (at least for the first two families) we are dealing with all
the quarks and leptons one is accustomed to within the set $(U_{1,2},D_{1,2},
N_{1,2},L_{1,2})$. The only care needed is to ensure that
left and right fields are distinguished by their isospin and hypercharge 
assignments. There are two ways to do this: one can either assume that the
zeta coordinates for matter are specifically tied to full Dirac fields 
$\Psi$ and introduce appropriate chirality projection operators $P_\pm =
(1\pm i\gamma_5)/2$, or one can presume that $\Psi$ is a left-handed to begin with
\footnote{But remember that fields reflected about the diagonal contain
the left-handed antiparticles so we are in effect including the right-handed
particle states too.} and associate generators involving $\zeta$ with left
chirality but those involving $\bar{\zeta}$ with right chirality.
In the first approach for gauging the standard SU(3)$\times$SU(2)$_L \times$U(1)
we would construct the covariant derivative acting on Dirac $\Psi$:
\begin{equation}
\gamma.D \equiv \gamma.[\partial -ig{\rm W}P_- -ig'B(Y_-P_- + Y_+P_+)]
\end{equation}
where
$$Y_- = -(\zeta^0\partial_0+\bar{\zeta}_0\bar{\partial}^0 +
            \zeta^4\partial_4+\bar{\zeta}_4\bar{\partial}^4)/2,\quad
  Y_+ = -(\zeta^4\partial_4+\bar{\zeta}_4\bar{\partial}^4).$$
In the second approach where $\psi$ is left-handed, we would take only one half of
the isospin operators, namely ${\rm T}_+ =\zeta^0\partial_4,\,
{\rm T}_-=\zeta^4\partial_0, {\rm T}_3 =(\zeta^0\partial_0 - 
\zeta^4\partial_4)/2$  and correspondingly ${\rm Y}
 = -(\zeta^0\partial_0+\zeta^4\partial_4)/2-\bar{\zeta}_4\bar{\partial}^4$.
(This guarantees that the right-handed particle fields are automatically weak 
isosinglets and is more elegant than the first approach.) Anyhow, with this 
halved set of generators of the second approach acting on $\Psi_L$ the 
covariant derivative reads
\begin{equation}
\gamma.D = \gamma.[\partial - ig{\rm W} - ig'B{\rm Y}]
\end{equation}

These manoeuvres just beg the question as to why one cannot gauge a larger
set of group generators. Of course one can, like any unified gauge model, but
unfortunately nature has picked on the standard model subset, the vagary of
picking on left-handed isospin being particularly quirky. Ultimately our
anticommuting coordinate scheme invites us to gauge the full Sp(10) group for 
the 10 real anticommuting properties, but then we would be hard-pressed to
explain where lie all the other gauge fields, not to mention the problem of
handling their leptoquark components. Of course this difficulty is common
to all unified schemes; the only virtue of ours is that the extra particles
and states are severely circumscribed and from the point of view of `economy'
this is surely a distinct advantage; also we get a concrete realization of
the group structure.

One last point concerns gauge anomalies. It is clear that for the lowest two
families, which are common to our approach and the conventional one, there can 
be none, i.e. there is complete cancellation between leptons and quarks,
per generation. Although we have not investigated what happens to the third 
generation which contains $U,D,X$ in a triplet and must be considered with
some of the other states, it is fairly clear that the anomalies will
disappear again for the standard model gauging because SU(3) is vectorial, SU(2)
has no anomalies and the trace of the charge operator vanishes. As far as the
unified gauge groups SU(5) or Sp(10) are concerned it would be
surprising if the same thing did not happen again, as this scheme has a lot
in common with SO(10) where we know there are no anomalies; for example the first
row of the magic square is simply a 16-fold of SO(10); however we
readily admit that the matter has not been defintively settled.

Finally, with regard to the Lagrangian for the gauge field, this can be
obtained in the same way through the extended space-time-property metric, outlined
in an earlier paper \cite{RD} for QED, but neglecting the gravitational part.
It falls out of course as the non-abelian version of the Maxwell Lagrangian.

\section{Flavour mixing and mass matrices}
The coupling of the superfield $\Psi$ to the vacuum Hiigs superfield $\langle
\Phi\rangle$ will induce flavour mixing. This arises via the interaction
$4\int d^5\zeta d^5\bar{\zeta}\,\,\bar{\Psi}\langle\Phi\rangle\Psi$, where we
ignore an overall coupling constant for the purposes of this work (absorbing it
in $\Phi$). Based on the {\em assumption} that the Higgs field is anti-selfdual
like $\Psi$, only nine expectation values make an appearance, labelled 
${\cal M}, {\cal A},\ldots {\cal H}$ of which just the final one is complex.
(Bear in mind that without antiduality there would be twice as many independent 
expectation values.) To see how to arrive at the results below we give an example, 
the flavour mixing of the first two charged leptons with one another, ignoring 
other components. First form the product $\bar{\Psi}\Psi$ and pick out the 
relevant parts, viz.
\begin{eqnarray*}
2\bar{\Psi}\Psi&\supset&\bar{L}_1L_1\bar{\zeta}_4\zeta^4(1-2\bar{\zeta}_0\zeta^0
  \bar{\zeta}_1\zeta^1\bar{\zeta}_2\zeta^2\bar{\zeta}_3\zeta^3) +\\
& & \bar{L}_2L_2\bar{\zeta}_4\zeta^4(-2\bar{\zeta}_0\zeta^0
  \bar{\zeta}_1\zeta^1\bar{\zeta}_2\zeta^2\bar{\zeta}_3\zeta^3) +\\ 
& & (\bar{L}_1L_2+\bar{L}_2L_1)\bar{\zeta}_4\zeta^4(\bar{\zeta}_0\zeta^0-
  \bar{\zeta}_1\zeta^1\bar{\zeta}_2\zeta^2\bar{\zeta}_3\zeta^3)+\cdots
\end{eqnarray*}
Then multiply by $\langle\Phi\rangle$ and pick out the zeta-product factor
$(\bar{\zeta}_\mu\zeta^\mu)^5/5!$, needed for Berezin integration. One obtains
the terms
$$4\int\!d^5\zeta d^5\bar{\zeta}\,\,\bar{\Psi}\langle\Phi\rangle\Psi\supset
  (2{\cal M}+{\cal B})\,\bar{L}_1L_1+2{\cal M}\,\bar{L}_2L_2+
  ({\cal A}+{\cal D})(\bar{L}_1L_2+\bar{L}_2L_1)+\cdots $$
and so on. Proceeding in this way, we can list the entire set of flavour 
mixing mass matrices $M_{i,j}$ that arise, where the range of indices,
$i,j$ depends on the particle type. For charged leptons the hermitian matrix 
$M$ reads
$$2M(L) \rightarrow 
\left( \begin{array}{cccccc}
       2{\cal M}+{\cal B} & {\cal A}+{\cal D} & {\cal C}+{\cal F} &
       {\cal E}-{\cal G} & -{\cal H}^* & {\cal H}^* \\
       {\cal A}+{\cal D} & 2{\cal M} & -{\cal G} & {\cal C} &{\cal H}^* & 0 \\
       {\cal C}+{\cal F} & -{\cal G} & 2({\cal M}-{\cal E}/\sqrt{3}) &
       {\cal A}-2{\cal C}/\sqrt{3} & 0 & 0\\
       {\cal E}-{\cal G} & {\cal C} & {\cal A}-2{\cal C}/\sqrt{3} & 
       2{\cal M} & 0 & 0 \\
       -{\cal H} & {\cal H} & 0 & 0 & 2{\cal M}+{\cal D} & {\cal A}+{\cal B}\\
       {\cal H} & 0 & 0 & 0 & {\cal A}+{\cal B} & 2{\cal M}
  \end{array} \right).$$
The neutrinos' matrix is smaller (just $4\times 4$):
$$2M(N) \rightarrow 
\left( \begin{array}{cccc}
   2{\cal M}+{\cal A}&{\cal B}+{\cal D}&{\cal C}+{\cal E}&{\cal F}-{\cal G}\\
   {\cal B}+{\cal D} & 2{\cal M} & -{\cal G} & {\cal C} \\
   {\cal C}+{\cal E} & -{\cal G} & 2({\cal M}-{\cal F}/\sqrt{3}) &
   {\cal B}-2{\cal C}/\sqrt{3} \\
   {\cal F}-{\cal G} & {\cal C} & {\cal B}-2{\cal C}/\sqrt{3} & 2{\cal M}
 \end{array} \right),$$
as is that of the $U$-quarks:
$$2M(U) \rightarrow 
\left( \begin{array}{cccc}
   2{\cal M}+{\cal F}/\sqrt{3}&{\cal B}+{\cal C}/\sqrt{3}&-{\cal H}^* & 0\\
   {\cal B}+{\cal C}/\sqrt{3} & 2{\cal M} & 0 & 0 \\
   -{\cal H} & 0 & 2{\cal M}+{\cal G}/\sqrt{3} & 2{\cal C}/\sqrt{3} \\
   0 & 0  & 2{\cal C}/\sqrt{3} & 2{\cal M}
 \end{array} \right),$$
but the $D$-quarks' $2M(D)$ matrix is very complicated, being $8\times 8$:
$$\left( \begin{array}{cccccccc}
       2{\cal M}\!+\!\frac{{\cal E}}{\sqrt{3}}&{\cal A}\!+\!\!
       \frac{{\cal C}}{\sqrt{3}}&
       -{\cal H}^*& 0 &{\cal H}^* & -{\cal H}^* & 0 & 0\\
       {\cal A}\!+\!\frac{{\cal C}}{\sqrt{3}} & 2{\cal M}& 0 & 0 &
       {\cal H}^* & 0 & 0 & 0 \\
       -{\cal H}&0&2{\cal M}\!+\!\frac{\cal G}{\sqrt{3}}&\frac{2{\cal C}}{\sqrt{3}}&
       \frac{{\cal E}-{\cal F}}{\sqrt{6}}\!+\!\frac{{\cal A}-{\cal B}}{\sqrt{2}}&0& 
       \frac{{\cal E}-{\cal F}}{\sqrt{3}}&\frac{{\cal B}-{\cal A}}{\sqrt{2}}\\       
       0&0& \frac{2{\cal C}}{\sqrt{3}} & 2{\cal M} & 
       \frac{2({\cal F}-{\cal E})}{\sqrt{6}}& 0 & {\cal A}\!-\!{\cal B} & 0 \\
       {\cal H} & {\cal H} & \frac{{\cal E}-{\cal F}}{\sqrt{6}}\!+\!
       \frac{{\cal A}-{\cal B}}{\sqrt{2}}&\frac{2({\cal F}-{\cal E})}{\sqrt{6}}&
       2{\cal M}\!+\!\frac{{\cal C}}{\sqrt{3}}&\frac{{\cal E}+{\cal F}}{\sqrt{6}}
       \!+\!\frac{{\cal A}+{\cal B}}{\sqrt{2}}&\frac{2({\cal G}+{\cal C})}
       {\sqrt{3}}&{\cal D}\!-\!\frac{{\cal G}}{\sqrt{3}} \\
       -{\cal H} & 0 & 0 & 0 & 
       \frac{{\cal E}+{\cal F}}{\sqrt{6}}\!+\!\frac{{\cal A}+{\cal B}}{\sqrt{2}} &
       2{\cal M} & -\frac{{\cal E} + {\cal F}}{\sqrt{3}} &
       \frac{{\cal A}+{\cal B}}{\sqrt{2}}\\
       0 & 0 & \frac{{\cal E}-{\cal F}}{\sqrt{3}} & {\cal A}-{\cal B} &
       \frac{2({\cal G}+{\cal C})}{\sqrt{3}}&-\frac{{\cal E}+{\cal F}}{\sqrt{3}}
       & 2{\cal M}\!-\!{\cal D} & -\frac{2{\cal C}}{\sqrt{6}}\\
       0 & 0 & \frac{{\cal B}-{\cal A}}{\sqrt{2}} & 0 &
       {\cal D}\!-\!\frac{{\cal G}}{\sqrt{3}} & \frac{{\cal A}+{\cal B}}{\sqrt{2}}&
       -\frac{2{\cal C}}{\sqrt{6}} & 2{\cal M}
 \end{array} \right).$$

These flavour matrices may be diagonalised by unitary tranformations $V$ in the
usual way: $m(u) = V(U)M(U)V^\dag(U)$, etc. And of course $V_{UD}\equiv
V^\dag(U)V(D)$ is also unitary in the standard picture because all quarks
are weak isodoublets. But this is not what occurs in our scheme; to see what
happens, focus on the charged weak interactions mediated by $W^\pm$. Realizing
that its coupling to the weak isotriplets differ by a factor of $\sqrt{2}$ from
those of the isodoublets, we meet the terms
$$W^+[(\bar{U}_1D_1+\bar{U}_2D_2) +\sqrt{2}(\bar{U}_3D_3+\bar{U}_4D_4
 + \bar{D}_3X_3+\bar{D}_4X_4)] + {\rm ~h.c}$$
which transcribe to the mass eigenvectors $\upsilon,\delta$:
$$W^+\bar{\upsilon}_j[(V_{1j}^*(U)V_{1k}(D)+V_{2j}^*(U)V_{2k}(D))
 +\sqrt{2}(V_{3j}^*(U)V_{4k}(D)+V_{4j}^*(U)V_{4k}(D))]\delta_k$$
$$ + {\rm h.c.} + WDX{\rm ~interaction-terms}.$$
(Here the first three $\upsilon$ correspond to $c,u,t$ quarks and the
first three $\delta$ to $s,d,b$ quarks.)
No longer is the effective $V_{UD}$ matrix unitary. Depending on the amount of 
flavour mixing of weak isodoublets with isotriplets and isosinglets $D_{5,..8}$,
we anticipate that small departures from unitarity will arise; also weak 
interactions of $b$ and $t$ quarks will be somewhat larger than those of the first and
second generations. The same remarks apply to the leptons in a modified manner:
while the first four generations of $N,L$ are weak isodoublets, $L_{5,6}$ are
isosinglets and it is their flavour mixing with the lighter $L$s that may
may induce a departure from unitarity of the 4x4 $V_{LN}$ matrix.

We are ready to make a stab at numerical values of masses and flavour mixings
more in an attempt to understand the mechanics than in trying to accurately
reproduce known mixings. It is futile to aim at exact figures for the known
twelve lepton and quark masses plus all their CKM-type mixings with only nine
parameters, $\cal A$ to $\cal H$, for we are looking at figures which range from
about $10^{-2}$ eV (for neutrinos) to $10^{11}$ eV (for the top quark), in other 
words thirteen orders of magnitude, with many masses interspersed in that range.
It is more likely that we can achieve a fit to data with a Higgs superfield that 
is not self-dual as that doubles the numbers of degrees of fitting freedom,
but we leave that to future work. For the present we shall simply ignore the
leptons and stick with the four $U$ quarks and eight $D$ quarks as getting
even these to be roughly right represents a modicum of progress.

We have not searched through the whole of parameter space in presenting
the following set of values; rather our aim has been to get acceptable values 
for $m_u,m_d,m_c,m_s,m_t,m_b$ and ensure that the other quark masses are
heavy, in regions not yer fully explored. The hope is that this can yield
flavour mixing matrices \footnote{We have not bothered to include a phase
in the complex field $\cal H$ which is the source of CP-violation as it 
introduces another layer of complexity.} that are not outright silly. Here is 
what we have adopted in units of MeV:
$${\cal M}=3, {\cal C}=-168000, {\cal B}=78+{\cal C}/\sqrt{3}, 
 {\cal A} = 5 - {\cal C}/\sqrt{3},$$
$${\cal D}=7700000, {\cal E}=600, {\cal F}=2600, {\cal G}=70000, 
 {\cal H} = 100$$
and thereby obtained
$$ m_u = 2, m_c = 1510, m_t = -175000, m_d = 6, m_s = 354, m_b = -4990.$$
[Negative mass eigenvalues are not particularly worrying since they can be
reversed by a $\gamma_5$ transformation on the eigenfields or equivalently
a relative change in phase of the left and right handed chiral components.]
These numbers are satisfactory and so too are the masses of the other quarks
which lie in the range above 175 GeV. (Actually there are strong indications
from the analysis that another D-quark may exist almost degenerate with the
top.)

Less successful is our determination of the mixing angles; the mass diagonalised 
states of $U$-quarks are not too bad, being given by the V-matrix:
\begin{equation}
\left(\!\!\begin{array}{c} c \\ u \\ t \\ \upsilon_4\\.\end{array}\!\! \right) =
\left( \begin{array}{ccccc}0.9987&0.052&4\times 10^{-6}&5\!\times\!10^{-4} & .\\
   -0.052& 0.9987 & 6\!\times\! 10^{-10} & 3 \!\times\! 10^{-4} & . \\
   3.8\!\times\! 10^{-4} & -1.7\!\times\! 10^{-7} & 0.67 & 0.74 & . \\
   -3.4\!\times\! 10^{-4} & -10^{-7} & 0.74 & -0.67 & .\\
  . & .& . & .& . \end{array}\right)
\left( \begin{array}{c} U_1 \\ U_2 \\ U_3 \\ U_4 \\ . \end{array}\right) ,
\end{equation}
but the $D$-quarks' V-matrix is more problematic:
\begin{equation}
\left(\!\!\begin{array}{c} s \\ d \\ b \\ \delta_4\\.\end{array}\!\! \right) =
\left( \begin{array}{ccccc}0.9997&0.014&4\times 10^{-6}&-5\!\times\!10^{-4} & .\\
   -0.014& 0.9999 & 6\!\times\! 10^{-8} & 7\!\times\! 10^{-6} & . \\
   0.02 & 0.00034 &-1.5\times 10^{-4}& 3.8\times 10^{-5} & . \\
   -3.8\!\times\! 10^{-4} &  -10^{-7}&-0.67 & -0.74  & .\\
  . & .& . & .& . \end{array}\!\!\right)
\left(\!\! \begin{array}{c} D_1 \\ D_2 \\ D_3 \\ D_4 \\ . \end{array}\!\!\right) ,
\end{equation}
because it means that the $b$ quark is largely $D_6$ (an electroweak singlet!)
which is very far from conventional wisdom. This is a definite failure of our
parameter choice. Given that we have abandoned any attempt at fitting the
(light) lepton masses, as is clear from the scales above, we will not
apologize any more. If one were to double the Higgs field expectation values
by discarding selfduality we would be in better shape all round, but that
would require a search through an 18-parameter space which is non-trivial!

In summary we reiterate the places where we differ from the standard model.
\begin{itemize}
\item We have at least four generations of quarks and leptons. Of the six 
charged leptons two are weak isosinglets, and of the eight $D$-quarks four
are weak isosinglets.
\item The top and bottom quarks belong to a weak isotriplet, together with 
another quark which we have designated $X$.
\item The flavour mixing induced by the super-Higgs components can result
in a $V_{UD}$ which is not quite unitary because the third generation is
an isotriplet; this would be a good experimental test of our scheme.
\item Colour sextet quarks with $F=1/3$, $Q=-1/3$ arise in property
combinations $(\bar{\zeta})^2\zeta$ where the zeta only carry chromicity.
If baryons carry small admixtures of these sextets in conjunction with
ordinary $u,d$ quarks this might help explain features of newly discovered
narrow width resonances that are interpreted as multiquarks.
\end{itemize}

\section*{Appendix - Normalized superfields}

Here we list the anti-selfdual components of the superfields $\Psi$ and $\Phi$,
arranged so that after Berezin integration they are properly normalized; our
convention is that $\int d^5\zeta d^5\bar{\zeta}\,(\bar{\zeta}_\mu\zeta^\mu)
\equiv -5!$\,\,We only list relevant parts that contain leptons and colour
triplet quarks, these being fields of primary interest, and start with red 
$U$-quarks; the latter are associated with (red) label 1 in the $\zeta$ 
expansion---not to be confused with the flavour counting subscripts carried by 
$U$:
\begin{eqnarray}
2\Psi_U \supset
2\Psi_{U\,{\rm red}}&=&[\bar{\zeta}_3\bar{\zeta}_2\bar{\zeta}_0\,U_1 + 
  U_1^c\,\zeta^0\zeta^2\zeta^3](1+\bar{\zeta}_4\zeta^4\bar{\zeta}_1\zeta^1) +
 \nonumber\\
 & & [\bar{\zeta}_3\bar{\zeta}_2\bar{\zeta}_0\,U_2 + 
  U_2^c\,\zeta^0\zeta^2\zeta^3](\bar{\zeta}_1\zeta^1+\bar{\zeta}_4\zeta^4) +
 \nonumber \\
 & &[\bar{\zeta}_0\zeta^1\zeta^4\,U_3+U_3^c\,\bar{\zeta}_4\bar{\zeta}_1\zeta^0]
  (1+\bar{\zeta}_2\zeta^2\bar{\zeta}_3\zeta^3) + \nonumber \\
 & &[\bar{\zeta}_0\zeta^1\zeta^4\,U_4+U_4^c\,\bar{\zeta}_4\bar{\zeta}_1\zeta^0]
  (\bar{\zeta}_2\zeta^2+\bar{\zeta}_3\zeta^3).
\end{eqnarray}
Because this is the first time we encounter it, let us also spell out the
adjoint components, taking due care with sign changes:
\begin{eqnarray}
2\bar{\Psi}_U \supset
2\bar{\Psi}_{U\,{\rm red}}&=&[\bar{U}_1\,\zeta^0\zeta^2\zeta^3 -
  \bar{\zeta}_3\bar{\zeta}_2\bar{\zeta}_0\,\bar{U_1^c}]
 (1+\bar{\zeta}_4\zeta^4\bar{\zeta}_1\zeta^1) + \nonumber\\
 & & [\bar{U}_2\,\zeta^0\zeta^2\zeta^3 -
  \bar{\zeta}_3\bar{\zeta}_2\bar{\zeta}_0\,\bar{U_2^c}]
 (\bar{\zeta}_1\zeta^1+\bar{\zeta}_4\zeta^4) + \nonumber \\
 & & [\bar{U_3}\,\bar{\zeta}_4\bar{\zeta}_1\zeta^0 -
  \bar{\zeta}_0\zeta^1\zeta^4\,\bar{U_3^c}]
  (1+\bar{\zeta}_2\zeta^2\bar{\zeta}_3\zeta^3) + \nonumber \\
 & &[\bar{U_4}\,\bar{\zeta}_4\bar{\zeta}_1\zeta^0 -
  \bar{\zeta}_0\zeta^1\zeta^4\,\bar{U_4^c}]
  (\bar{\zeta}_2\zeta^2+\bar{\zeta}_3\zeta^3).
\end{eqnarray}
One may readily check that $\int\!d^5\zeta d^5\bar{\zeta}\,
 (\bar{\Psi}_{U\,{\rm red}}\Psi_{U\,{\rm red}})=(\bar{U}_1U_1+\bar{U}_2U_2+
 \bar{U}_3U_3+\bar{U}_4U_4)$. 

Next we list the red down-type quark field components, again designated by
(red) label 1 so far as the $\zeta$ coordinates are concerned. It contains
twice as many pieces as the up-type quarks:
\begin{eqnarray}
2\bar{\Psi}_D \supset
2\bar{\Psi}_{D\,{\rm red}}&=&[\bar{\zeta}_3\bar{\zeta}_2\bar{\zeta}_4\,D_1
  + D_1^c\,\zeta^4\zeta^2\zeta^3](1+\bar{\zeta}_0\zeta^0\bar{\zeta}_1\zeta^1) 
 + \nonumber\\
 & & [\bar{\zeta}_3\bar{\zeta}_2\bar{\zeta}_4\,D_2
 + D_2^c\,\zeta^4\zeta^2\zeta^3](\bar{\zeta}_1\zeta^1+\bar{\zeta}_0\zeta^0) 
 + \nonumber \\
 & & [\bar{\zeta}_1\,D_3^c+D_3\,\zeta^1](\bar{\zeta}_0\zeta^0-
  \bar{\zeta}_4\zeta^4)(1+\bar{\zeta}_2\zeta^2\bar{\zeta}_3\zeta^3)/\sqrt{2}
 + \nonumber\\
& & [\bar{\zeta}_1\,D_4^c+D_4\,\zeta^1](\bar{\zeta}_2\zeta^2+
  \bar{\zeta}_3\zeta^3)(\bar{\zeta}_0\zeta^0-\bar{\zeta}_4\zeta^4)/\sqrt{2}
 + \nonumber \\
 & & [\bar{\zeta}_1\,D_5^c+D_5\,\zeta^1](1 - 
 \bar{\zeta}_2\zeta^2\bar{\zeta}_3\zeta^3\bar{\zeta}_0\zeta^0\bar{\zeta}_4
  \zeta^4) + \nonumber\\
 & & [\bar{\zeta}_1\,D_6^c+D_6\,\zeta^1](\bar{\zeta}_0\zeta^0+
  \bar{\zeta}_4\zeta^4)(1-\bar{\zeta}_2\zeta^2\bar{\zeta}_3\zeta^3)/\sqrt{2}
 + \nonumber\\
& & [\bar{\zeta}_1\,D_7^c+D_7\,\zeta^1](\bar{\zeta}_2\zeta^2+
  \bar{\zeta}_3\zeta^3)(1-\bar{\zeta}_0\zeta^0\bar{\zeta}_4\zeta^4)/\sqrt{2}
 + \nonumber \\
 & & [\bar{\zeta}_1\,D_8^c+D_8\,\zeta^1](\bar{\zeta}_0\zeta^0
 \bar{\zeta}_4\zeta^4-\bar{\zeta}_2\zeta^2\bar{\zeta}_3\zeta^3).
\end{eqnarray}
Note that the last four components are weak isosinglets. Here again, 
 $\int\!d^5\zeta d^5\bar{\zeta}\,
 (\bar{\Psi}_{D\,{\rm red}}\Psi_{D\,{\rm red}})=(\bar{D}_1D_1+\bar{D}_2D_2+
 \bar{D}_3D_3+\bar{D}_4D_4+\bar{D}_5D_5+\bar{D}_6D_6+\bar{D}_7D_7+
 \bar{D}_8D_8)$.

Turning to leptons, begin with the four neutrinos contained in
\begin{eqnarray}
2\Psi_N &=&[\bar{\zeta}_0\,N_1+N_1^c\,\zeta^0]
 (1-\bar{\zeta}_4\zeta^4\bar{\zeta}_1\zeta^1\bar{\zeta}_2\zeta^2\bar{\zeta}_3
 \zeta^3) + \nonumber\\
 & & [\bar{\zeta}_0\,N_2+N_2^c\,\zeta^0](\bar{\zeta}_4\zeta^4-
  \bar{\zeta}_1\zeta^1\bar{\zeta}_2\zeta^2\bar{\zeta}_3\zeta^3) + \nonumber \\
&&[\bar{\zeta}_0\,N_3+N_3^c\,\zeta^0](\bar{\zeta}_i\zeta^i-\bar{\zeta}_4\zeta^4
 \bar{\zeta}_j\zeta^j\bar{\zeta}_k\zeta^k/2)/\sqrt{3} + \nonumber \\
& &[\bar{\zeta}_0\,N_4+N_4^c\,\zeta^0](\bar{\zeta}_i\zeta^i\bar{\zeta}_4\zeta^4
 - \bar{\zeta}_j\zeta^j\bar{\zeta}_k\zeta^k/2)/\sqrt{3}.
\end{eqnarray}
and then write out the charged leptons (very similar except for an extra two 
entries that are weak isospin singlets):
\begin{eqnarray}
2\Psi_L &=&[\bar{\zeta}_4\,L_1+L_1^c\,\zeta^4]
 (1-\bar{\zeta}_0\zeta^0\bar{\zeta}_1\zeta^1\bar{\zeta}_2\zeta^2\bar{\zeta}_3
 \zeta^3) + \nonumber\\
 & & [\bar{\zeta}_4\,L_2+L_2^c\,\zeta^4](\bar{\zeta}_0\zeta^0-
  \bar{\zeta}_1\zeta^1\bar{\zeta}_2\zeta^2\bar{\zeta}_3\zeta^3) + \nonumber \\
&&[\bar{\zeta}_0\,L_3+L_3^c\,\zeta^0](\bar{\zeta}_i\zeta^i-\bar{\zeta}_0\zeta^0
 \bar{\zeta}_j\zeta^j\bar{\zeta}_k\zeta^k/2)/\sqrt{3} + \nonumber \\
& &[\bar{\zeta}_4\,L_4+L_4^c\,\zeta^4](\bar{\zeta}_i\zeta^i\bar{\zeta}_0\zeta^0
 - \bar{\zeta}_j\zeta^j\bar{\zeta}_k\zeta^k/2)/\sqrt{3} + \nonumber \\
& &[\bar{\zeta}_3\bar{\zeta}_2\bar{\zeta}_1\,L_5^c
  + L_5\,\zeta^1\zeta^2\zeta^3](1+\bar{\zeta}_0\zeta^0\bar{\zeta}_4\zeta^4) 
 + \nonumber\\
 & & [\bar{\zeta}_3\bar{\zeta}_2\bar{\zeta}_1\,L_6^c
 + L_6\,\zeta^1\zeta^2\zeta^3](\bar{\zeta}_0\zeta^0+\bar{\zeta}_4\zeta^4). 
\end{eqnarray}
Altogether one may verify that
$\int\!d^5\zeta d^5\bar{\zeta}\,
 (\bar{\Psi}_N\Psi_N +\bar{\Psi}_L\Psi_L)=(\bar{N}_1N_1+\bar{N}_2N_2+
 \bar{N}_3N_3+\bar{N}_4N_4+\bar{L}_1L_1+\bar{L}_2L_2+
 \bar{L}_3L_3+\bar{L}_4L_4+\bar{L}_5L_5+\bar{L}_6L_6)$, which is 
perfectly satisfactory.

Finally we spell out the parts of the Higgs superfield that have $Q=F=0$.
Having {\em assumed} anti-selfduality these components read:
\begin{eqnarray}
 2\langle\Phi\rangle &=&{\cal M}(1-\bar{\zeta}_0\zeta^0\bar{\zeta}_1\zeta^1
 \bar{\zeta}_2\zeta^2\bar{\zeta}_3\zeta^3\bar{\zeta}_4\zeta^4)+ 
 {\cal A}(\bar{\zeta}_0\zeta^0-\bar{\zeta}_1\zeta^1
 \bar{\zeta}_2\zeta^2\bar{\zeta}_3\zeta^3\bar{\zeta}_4\zeta^4)+ \nonumber \\
 & &{\cal B}(\bar{\zeta}_4\zeta^4-\bar{\zeta}_1\zeta^1
 \bar{\zeta}_2\zeta^2\bar{\zeta}_3\zeta^3\bar{\zeta}_0\zeta^0)+
 {\cal C}(\bar{\zeta}_i\zeta^i-\bar{\zeta}_0\zeta^0
 \bar{\zeta}_4\zeta^4\bar{\zeta}_j\zeta^j\bar{\zeta}_k\zeta^k/2)/\sqrt{3}
 + \nonumber \\
& &{\cal D}(\bar{\zeta}_0\zeta^0\bar{\zeta}_4\zeta^4 -
 \bar{\zeta}_1\zeta^1\bar{\zeta}_2\zeta^2\bar{\zeta}_3\zeta^3) +
 {\cal E}(\bar{\zeta}_i\zeta^i\bar{\zeta}_0\zeta^0-
 \bar{\zeta}_4\zeta^4\bar{\zeta}_j\zeta^j\bar{\zeta}_k\zeta^k/2)/\sqrt{3}
 + \nonumber \\
& &{\cal F}(\bar{\zeta}_i\zeta^i\bar{\zeta}_4\zeta^4-
 \bar{\zeta}_0\zeta^0\bar{\zeta}_j\zeta^j\bar{\zeta}_k\zeta^k/2)/\sqrt{3}
 + \nonumber \\
 & &{\cal G}(\bar{\zeta}_i\zeta^i\bar{\zeta}_j\zeta^j/2-
 \bar{\zeta}_4\zeta^4\bar{\zeta}_0\zeta^0\bar{\zeta}_k\zeta^k)/\sqrt{3} 
 + \nonumber \\
& & [{\cal H}\,\zeta^1\zeta^2\zeta^3\zeta^4 +
  {\cal H}^*\,\bar{\zeta}_4\bar{\zeta}_3\bar{\zeta}_2\bar{\zeta}_1]
   (1-\bar{\zeta}_0\zeta^0).
\end{eqnarray}
Except for $\cal H$ all other expectation values are real. In fact it is the 
$\cal H$ components that are responsible for CP-violation in this scheme. 
Checking it all out,
$\int\!d^5\zeta d^5\bar{\zeta}\,
 2\langle \Phi \rangle^2={\cal M}^2 + {\cal A}^2 + {\cal B}^2 + {\cal C}^2
 + {\cal D}^2 + {\cal E}^2 + {\cal F}^2 + {\cal G}^2 + 2{\cal H}^*{\cal H}$.
We stress that relaxing selfduality for $\Phi$ will double the number of
independent expectation values and provide more freedom in the mass matrices.

\end{document}